\documentclass[11pt]{article}
\usepackage{latexsym}

\def\be{\begin{equation}}
\def\ee{\end{equation}}
\def\bea{\begin{eqnarray}}
\def\eea{\end{eqnarray}}
\def\pa{\partial}

\def\fn{\footnote}

\def\case#1/#2{\textstyle\frac{#1}{#2}}

\begin{document}
\begin{titlepage}

\vspace{.7in}

\begin{center}
\Large\bf
{STRONG-COUPLED RELATIVITY WITHOUT RELATIVITY}\normalfont
\\
\vspace{.7in} \normalsize \large{Edward Anderson}
\\
\normalsize
\vspace{.4in}
{\em  Astronomy Unit, School of Mathematical Sciences, \\
Queen Mary, Mile End Road, London E1 4NS, U.K. }

\vspace{.2in}
\end{center}
\vspace{.3in}
\baselineskip=24pt

\begin{abstract}

GR can be interpreted as a theory of evolving 3-geometries.  
A recent such formulation, the 3-space approach of Barbour, Foster and \'{O} Murchadha, 
also permits the construction of a limited number of other theories of evolving 3-geometries, 
including \it conformal gravity \normalfont and \it strong gravity\normalfont.  
In this paper, we use the 3-space approach to construct a 1-parameter family of theories which generalize strong gravity.
The usual strong gravity is the strong-coupled limit of GR, 
which is appropriate near singularities 
and is one of very few regimes of GR which is amenable to quantization.  
Our new strong gravity theories are similar limits of scalar-tensor theories such as Brans--Dicke theory, and are likewise appropriate near singularities.  
They represent an extension of the regime amenable to quantization, which furthermore spans two qualitatively different types of inner product.  

We find that these strong gravity theories permit coupling only to ultralocal matter fields and that they prevent gauge theory.  Thus in the classical picture, 
gauge theory breaks down (rather than undergoing unification) as one approaches the GR initial singularity.  

\end{abstract}

PACS  number 04.20.Fy

\vspace{.3in}
Electronic address: e.anderson@qmul.ac.uk

\end{titlepage}

\section{Introduction}

In the Arnowitt--Deser--Misner (ADM) \cite{ADM} formulation of GR, the action 
is\fn{We use lower-case Latin letters for 3-space indices.  Round brackets denote symmetrization of indices, and square brackets denote antisymmetrization. 
Indices unaffected by the (anti)symmetrization are set between vertical lines. $g_{ij}$ is the spatial 3-metric with determinant $g$ and conjugate 
momentum $p^{ij}$.  $R$ is the spatial Ricci scalar and spatial covariant derivatives are denoted by a $\nabla$ or a semicolon.} 
\be
S = \int\textrm{d}\lambda\int\textrm{d}^3x(p^{ij}\dot{g}_{ij} - \xi^i{\cal H}_i - N{\cal H} ), 
\label{SADM} 
\ee  
\be
{\cal H}_i \equiv = - 2\nabla_j{p_i}^j = 0, 
\label{mom}
\ee
\be
{\cal H} \equiv G_{ijkl}p^{ij}p^{kl} - \sqrt{g}R = 0,
\label{ham}
\ee
where $G_{ijkl} = \frac{1}{\sqrt{g}}\left(g_{i(k|}g_{j|l)} -\frac{1}{2}g_{ij}g_{kl}\right)$ is the DeWitt supermetric \cite{DeWitt}.  
The purpose of this formulation is to treat GR as a dynamical system.  The shift $\xi^i$ and lapse $N$ are merely auxiliary variables, 
variation with respect to which yields the momentum and Hamiltonian constraints (\ref{mom}) and (\ref{ham}).  
GR has 2 true dynamical variables per space point.  These are what is left of the 6 degrees of freedom of the 3-metric $g_{ij}$, once one has taken the 4 constraints into account.   
Thus a configuration space for GR, that has a fourfold redundancy per space point, is Riem: the space of 3-metrics on a fixed topology (taken here to be compact without boundary).   
It is relatively straightforward to understand the restriction placed on this by the momentum constraint, which generates the 3-diffeomorphisms: the true dynamical variables are 
contained within the 3 degrees of freedom of the 3-geometries and not among the 3 coordinates painted onto these 3-geometries.  Thus GR may be interpreted as a theory of 
evolving 3-geometries, or \it geometrodynamics \normalfont \cite{Battelle}.  The corresponding configuration space is the quotient space 
\be
\{\mbox{Superspace}\} = \frac{\{\mbox{Riem}\}}{\{\mbox{3-diffeomorphisms}\}}
\ee
which has a single redundancy per space point due to the still-remaining Hamiltonian constraint, ${\cal H}$, which then plays a central role in geometrodynamics.  

Possible classical and quantum interpretations of geometrodynamics have been discussed by Wheeler.  In particular \cite{Battelle}, he asked 
``If one did not know the Einstein--Hamilton--Jacobi equation, how might one hope to derive it straight off from plausible first principles without ever going through the 
formulation of the Einstein field equations themselves?". The first stage of answering this question involves finding a derivation for ${\cal H}$, which is the fully classical analogue of this 
equation.  Following Wheeler's suggestion of presupposing embeddability into a spacetime to answer the question, Hojman, Kucha\v{r} and Teitelboim (HKT) \cite{HKT} obtained the 
partial answer that the form (\ref{ham}) of ${\cal H}$ (with an additional optional cosmological constant term) is required in order for the constraints to close as the 
\it Dirac Algebra \normalfont, which is the condition to ensure this embeddability.  More recently, Barbour, Foster and \'{O} Murchadha (BF\'{O})'s distinct 3-space approach 
\cite{BOF} includes doing away with the presupposed embeddability by demanding mere closure, that is asking what self-consistent dynamical systems can describe evolving 
3-geometries.  Their first principles and method are as follows.  

BF\'{O}'s first principles are 1) to consider prospective laws of physics that are based on relative quantities alone, and 2) that there should be no overall notion of time.  
Principle 1) is to be implemented by working with actions on the appropriate relative configuration space (which is done indirectly, by the method of `best matching' 
\cite{BB} outlined below).  In working with configuration spaces, the whole system is represented therein by a single point, and the evolution of the system is the curve 
traced out by these points.  Because the \it whole \normalfont system is represented thus, there is nothing external to the system with respect to which the parametrization 
time-label $\lambda$ of the curve could be compared.  Thus the theory shoukd be invariant under an overall reparametrization of the time-label $\lambda$.  This is what is meant 
by principle 2) and its implementation by working with  reparametrization-invariant actions.  

To treat geometrodynamics, the relative configuration space in question is superspace.   In this context, \it best matching \normalfont is a method to implement 
3-diffeomorphism invariance, by correcting the bare velocities of all fields $\Psi$ present according to $\dot{\Psi} \longrightarrow \dot{\Psi} - \pounds_{\xi}\Psi$, where the 
dot is  $\frac{\pa}{\pa\lambda}$ and $\pounds_{\xi}$ denotes the Lie derivative with respect  to the vector field $\xi^i$.  This is an indirect implementation since nothing is 
done to eliminate any of the $g_{ij}$ and furthermore  $\xi_i$ has been introduced, so that one's action is on Riem $\times$ $\Xi$, where $\Xi$  is the space of the $\xi_i$.  But 
variation with respect to $\xi_i$ gives rise to the momentum constraint.  If one could solve this for $\xi_i$, then one could pass to superspace.  

The form of the GR action that Baierlein, Sharp and Wheeler (BSW) \cite{BSW} derived, 
\be
S_{\mbox{\scriptsize BSW\normalsize}}
= \int d\lambda \int d^3x\sqrt{g}\sqrt{RT_{\mbox{\scriptsize g\normalsize}}},
\ee 
\be
T_{\mbox{\scriptsize g\normalsize}} = \frac{1}{\sqrt{g}}G^{abcd}(\dot{g}_{ab} - \pounds_{\xi}g_{ab})( \dot{g}_{cd} - \pounds_{\xi}g_{cd})
\ee
may be taken to explicitly show that GR is indeed a theory of this form.  

Now, BF\'{O} systematically examined such reparametrization-invariant BSW-type actions 
\be
S_{\mbox{\scriptsize \normalsize}}
= \int d\lambda \int d^3x \sqrt{g}\mbox{\sffamily L\normalfont}(g_{ij}, \dot{g}_{ij}; \xi_i)
= \int d\lambda \int d^3x\sqrt{g}\sqrt{PT_{\mbox{\scriptsize W\normalsize}}},
\label{BSWac}
\ee
where the kinetic term  
\be
T_{\mbox{\scriptsize W\normalsize}} = \frac{1}{\sqrt{g}}G^{abcd}_{\mbox{\scriptsize W\normalsize}}
(\dot{g}_{ab} - \pounds_{\xi}g_{ab}) ( \dot{g}_{cd} - \pounds_{\xi}g_{cd}).
\label{TW}
\ee
is built using best matching and the most general \it ultralocal\normalfont\fn{By an expression being ultralocal, we mean that it does not contain spatial derivatives.}  
$\mbox{ }$ inverse supermetric, 

\noindent $G_{\mbox{\scriptsize W\normalsize}}^{ijkl} =  \sqrt{g}(g^{ik}g^{jl} - Wg^{ij}g^{kl})$, which is the inverse of the DeWitt supermetric when the  
free parameter $W$ takes the value $1$.  BF\'{O} chose a simple potential term, $P = \Lambda + sR$, for $\Lambda$ constant and without loss of generality $s \in \{1,0,-1\}$; 
furthermore they showed that some more complicated potentials failed to be consistent.  

BF\'{O} use Dirac's generalized Hamiltonian method \cite{Dirac} exhaustively \cite{AB}, as explained in Sec 2.  In outline, in addition to the ${\cal H}_i$ constraint (relation 
between the momenta) obtained by $\xi_i$-variation, a relation ${\cal H}_{\mbox{\scriptsize W\normalsize}}$ between the momenta arises purely from the local square-root form of 
the Lagrangian in (\ref{BSWac}).  Consistency and nontriviality then require ${\cal H}_{\mbox{\scriptsize W\normalsize}}$ and 
${\cal H}_i$ to be propagated by the equations of motion without the production of more constraints than the theory has remaining degrees of freedom.  
The remarkable consequences of this exhaustive interpretation include enforcing 
${\cal H}_{\mbox{\scriptsize W\normalsize}} =  {\cal H}_{({\mbox{\scriptsize W\normalsize}} =1)} \equiv {\cal H}$ in the Lorentzian ($s = 1$) case, as well as overruling the 
more complicated potentials, and giving further results on `adding on' general bosonic matter, to which we now turn.  

The reasons for `adding on' matter in answering Wheeler's question is that its 
context has changed, for his original hopes that vacuum geometrodynamics could be a unified theory (by extension of the Rainich--Misner--Wheeler `already-unified' theory of 
gravity and electromagnetism \cite{RMW} to include all the other fields of nature), have not been realized.  Thus already in 1980 Teitelboim \cite{Teitelboim} extended HKT's  
answer by `adding on' matter and this is also the way in which BF\'{O} have treated matter.  The BF\'{O} treatment appears to give some striking derivations of the classical 
laws of bosonic physics.  In particular, rather than being presupposed, both gauge theory and the light-cone structure for bosonic theories are enforced, and share a common 
origin in the propagation of ${\cal H}$.  Also in BF\'{O}'s treatment and its extension by the author and Barbour \cite{AB}, masslessness is enforced on fundamental vector 
fields; Maxwell and Yang--Mills theory are picked out.  

\mbox{ }

The 3-space approach is not just a reformulation of GR.  It is also a method by which a limited number of other theories of evolving 3-geometries may be constructed, which are 
very similar to the ADM formulation of GR.  One such example is \it conformal gravity \normalfont \cite{BO, CG}, which has some promising features as a realistic physical theory.  

Another example is the theory called \it{strong gravity} \normalfont since it corresponds to the strong-coupled limit of GR, in which Newton's gravitational constant 
$G \longrightarrow \infty $, or equivalently $c \longrightarrow 0$ \cite{STeitel}.  This is the opposite of the more common Galilean limit $c \longrightarrow \infty$ 
(see Fig. 1). In place of the Lorentz group and the light-cone structure, one has an ultralocal `Carroll group' \cite{STeitel} structure in which each point is entirely isolated 
from the others.  Strictly speaking, we must keep $\Lambda/G$ constant in evaluating this limit from GR.  Otherwise the theory is dynamically trivial.  

Strong gravity was first considered by Isham \cite{Isham} as a new regime about which one might construct a perturbative theory of quantum gravity, akin to Klauder's ultralocal 
field theory \cite{Klauderlit}.  In this paper, we consider strong gravity as a dynamically-consistent theory on its own merit.  Henneaux \cite{Henneaux} showed that it has an 
unusual 4-geometry resulting from the degeneracy of the 4-metric; one consequence of this is that, unlike in GR, the constraints and evolution equations do not combine to form a 
4-tensor equation.  Moreover, conformal gravity again does not appear to admit a clean interpretation as a 4-tensor theory.  So strong gravity may provide some intuition as to 
what is possible in such theories.

Strong gravity approximates GR near the cosmological singularity, making it a worthwhile regime to quantize \cite{Pilatilit}.  Strong gravity gives an independent Kasner 
universe at each spatial point, which is the conjectured behaviour of the general solution of GR.  Belinskii, Khalatnikov and Lifshitz conjectured mixmaster behaviour (a 
sequence of Kasner epochs) at each spatial point \cite{BKL} whilst straightforward Kasner behaviour at each spatial point can sometimes occur \cite{VDTandAVDT}.  There is 
growing numerical evidence for these behaviours \cite{BKLmod}.  The notion of strong gravity is related to the two most popular approaches to quantum gravity as follows.  
It is analogous to the tensionless string \cite{stringyrefs1}, and it admits an Ashtekar variable formulation \cite{bombelli, Husain} (see Sec 2.3). 

\mbox{ }

In Sec 2, we provide a further example: a 1-parameter family of theories of evolving 3-geometries.  Their discovery provides a \it different \normalfont answer to Wheeler's 
question from the uniqueness of BF\'{O}, in the case of the strong-coupled limit $s = 0$:  there is a consistent theory not only for the $W = 1$ DeWitt supermetric of the usual 
strong gravity, but also for any ultralocal invertible ($W \neq \frac{1}{3}$) supermetric.    Whereas in the GR case of the 3-space approach the presupposition or otherwise of 
best matching does not affect the final theory that arises by consistency, strong gravity illustrates that this presupposition can alter the final theory.  The spirit of the 
3-space approach is to treat our strong gravities as consistent theories of evolving 3-geometries in their own right, and hence akin to theories of gravity.  Furthermore we 
discuss how they can be related (as limits relevant to the very early universe) to the well-known scalar-tensor theories of gravity (Sec 2.1), and used as toy models toward 
the study of conformal gravity (Sec 2.2).    
 
GR offers only two regimes amenable to quantization: minisuperspace and strong gravity. This paper provides an enlargement of this second arena, for which different ranges of 
$W$ give rise to considerable mathematical differences \cite{Giulini}.  In particular, for $W < \frac{1}{3}$ one has theories with positive-definite inner products. The study 
of these could broaden the understanding of the inner product problem of quantum gravity \cite{Kuchar92}.  We furthermore discuss the possibility of the very early universe actually having a 
positive-definite inner product.  We finally show (Sec 2.3) that the Ashtekar variable formulation, of potentially great use in quantization, is not readily available for 
$W \neq 1$. 

In Secs 3.1 and 3.2, we couple to strong gravity scalar fields and many interacting vector fields respectively.  This enables fruitful comparison with matter coupling in the 
GR case (Secs 3.2 and 3.3), which leads to better understanding of some of the GR 3-space approach results.  First, we find that in BF\'{O}'s approach, strong gravity theories 
impose an ultralocal structure rather than a Lorentz one, and that they cause the breakdown of gauge theory, which reinforces BF\'{O}'s notion that the light-cone and gauge 
theory have a common origin in the 3-space approach to GR coupled with bosonic fields.  Second, we find that massive vector fields are readily included coupled to strong 
gravity.  This helps clarify the central role of the differential Gauss laws of electromagnetism and Yang--Mills theory in the masslessness of vector fields in the 3-space 
approach.   Since the first of these results includes the usual GR strong-coupled limit, it means that we have shown that gauge theory breaks down  
as one approaches the initial singularity of GR.

\section {Strong Gravity and the 3-Space Approach}

We are interested in finding consistent theories of evolving 3-geometries; 
we use BF\'{O}'s 3-space approach to construct them. 
Consider then the following best-matched, reparametrization-invariant action, 
\be
S_{\mbox{\scriptsize BSW \normalsize}}
= \int d\lambda \int d^3x \sqrt{g} \sqrt{s R + \Lambda}
\sqrt{T_{\mbox{\scriptsize W\normalsize}}}.
\label{ASBSW}
\ee
The supermetric whose inverse $G^{abcd}_{\mbox{\scriptsize W\normalsize}}$ is used in the construction of $T_{\mbox{\scriptsize W\normalsize}}$ is 
\be
G_{\mbox{\scriptsize X\normalsize}abcd} = \frac{1}{\sqrt{g}}\left(g_{ac}g_{bd} - \frac{X}{2}g_{ab}g_{cd}\right), 
\ee
for
\be
X = \frac{2W}{3W - 1},
\ee
so that the DeWitt case $W = 1$ corresponds to $X = 1$, and $W = 0$ corresponds to $X = 0$.   
$W = \frac{1}{3}$ is excluded from the treatment on grounds of noninvertibility.  

The canonical gravitational momenta are 
\be
p^{ij} \equiv \frac{\partial\mbox{\sffamily{L}\normalfont} }       { \partial\dot{g}_{ij}}
= G_{\mbox{\scriptsize W\normalsize}}^{ijcd}
\frac{1}{2N}(\dot{g}_{cd} - 2\nabla_{(c}\xi_{d)}),
\label{Sgmom}
\ee
where $2N =\sqrt{T_{\mbox{\scriptsize W\normalsize}}/(s R + \Lambda)}$.  We can invert 
(\ref{Sgmom}) to find an expression for $\dot{g}_{ij}$.  

The primary  Hamiltonian constraint follows directly from the local square-root form of the Lagrangian.    
Since this statement is important and recurrent in our work, we provide here its derivation and interpretation:   
\be
\begin{array}{l}
G_{\mbox{\scriptsize X\normalsize}ijkl}p^{ij}p^{kl}  =
G_{\mbox{\scriptsize X\normalsize}ijkl}
G_{\mbox{\scriptsize W\normalsize}}^{ijcd}\frac{1}{2N}(\dot{g}_{cd} - 2\nabla_{(c}\xi_{d)})
G_{\mbox{\scriptsize W\normalsize}}^{klab}\frac{1}{2N}(\dot{g}_{ab} - 2\nabla_{(a}\xi_{b)}) 
= \sqrt{g}\frac{T_{\mbox{\scriptsize W\normalsize}}}{(2N)^2} = \sqrt{g}(sR + \Lambda),
\end{array}
\ee
by (\ref{Sgmom}) and (\ref{TW}).  
Hence the momenta are not independent, but rather there is a relation between them, 
\be
{\cal H } \equiv \frac{1}{\sqrt{g}}\left(p^{ij}p_{ij} - \frac{X}{2}p^2\right) - \sqrt{g}(s R + \Lambda) = 0,
\label{GRHam}
\ee
where $p$ is the trace of $p^{ij}$.  
Such a relation between the momenta is called a constraint; it is called \it primary \normalfont to indicate that variation was not used in its derivation.  
Constraints arising from  variation are called \it secondary\normalfont; variation with respect to $\xi_i$ yields the secondary momentum constraint 
\be
{\cal H}_i \equiv -2\nabla_j{p_i}^j = 0.
\label{SGRMom}
\ee
If we can solve this as a differential equation for $\xi_i$ (which is the thin sandwich conjecture  \cite{TSlit}), then one is actually working on superspace.  

The Euler--Lagrange equations are 
\be
\begin{array}{ll}
\frac{\partial p^{ij}} {\partial \lambda} =
\frac{\delta\mbox{\sffamily\scriptsize L\normalfont\normalsize}}{\delta g_{ij}} =  & \sqrt{g}Ng^{ij}(s R + \Lambda) -\sqrt{g}sNR^{ij}
- \frac{2N}{\sqrt{g}}\left(p^{im}{p_m}^j -\frac{X}{2}p^{ij}p\right)
+ \sqrt{g}s(N^{;ij} - g^{ij}\nabla ^2 N) + \pounds_{\xi}p^{ij},
\end{array}
\ee
from which we evaluate $\dot{{\cal H}}$: 
\be
\begin{array}{c}
\frac{\partial}{\partial\lambda}
\left[ \frac{1}{\sqrt{g}}\left(p^{ij}p_{ij} - \frac{X}{2}p^2 \right) -\sqrt{g}(sR + \Lambda)\right]
= \frac{Np(3X - 2)}{2\sqrt{g}}
\left[\frac{1}{\sqrt{g}}\left(p^{ij}p_{ij} - \frac{X}{2}p^2 \right) -\sqrt{g}(sR + \Lambda)\right] \\
+ \pounds_{\xi}
\left[
\frac{1}{\sqrt{g}}
\left(
p^{ij}p_{ij} - \frac{X}{2} p^2 
\right) 
- \sqrt{g}(sR + \Lambda)
\right]
- \frac{2s} {N}(N^2{p^{ab}}_{;b})_{;a} + \frac{2s}{N}(1 - X)
\left(
N^2 p_{;a}
\right)
^{;a}.
\label{Sevolham}
\end{array}
\ee
We demand that this vanishes weakly in the sense of Dirac \cite{Dirac}, i.e. that it is zero modulo the constraints hitherto found, which we denote by the symbol $\approx$:  
$\dot{{\cal H}} \approx 0$. 
Notice that expression (\ref{Sevolham}) has been grouped 
so that the first two terms vanish by virtue of the Hamiltonian constraint and the third 
vanishes by virtue of the momentum constraint, which is thus an integrability condition for 
the Hamiltonian constraint in GR.  Thus we are left with just the last term.  

Supposing that this does not automatically vanish, then we would require new constraints.  But we can apply exhaustive arguments as spelled out in general in \cite{AB},
which rather cut down on how many further constraints can arise, since these quickly use up the theory's degrees of freedom.

There are three possibilities for (\ref{Sevolham}) to be consistent. 
First, $p_{;a} = 0$, which gives a further constraint $p/\sqrt{g} = const$. 
Then we require furthermore that the new constraint propagates. This 
gives additionally the constant mean curvature (CMC) slicing equation 
\be
\frac{\pa}{\pa\lambda}\left(\frac{p}{\sqrt{g}}\right) = C =  3\Lambda N + 2s(NR - \nabla^2N) + \frac{(3X - 2)Np^2}{2g},  
\label{slicingeq}
\ee
(for $C$ a spatial constant) which nontrivially fixes the lapse $N$ provided that $s \neq 0$.  This demonstrates that arbitrary geometries cannot be connected, 
which for $s \neq 0$ forces us to take the second possibility: that $X$ takes its DeWitt value $X = 1$.  This is the main result, `Relativity Without Relativity', 
of BF\'{O}'s paper, whereby GR is recovered.   Giulini had also noted that $X = 1$ is mathematically special \cite{Giulini}.

But clearly $s = 0$ gives a third possibility, for which any ultralocal supermetric is allowed.  The theory traditionally called strong gravity has $W = 1$ because it is 
obtained as a truncation of GR.  But our first result is that as far as dynamical consistency is concerned, we have now shown that there exists a family of such theories 
parametrized by $W$.  Because $W = 1/3$ is badly behaved, the family of dynamically-consistent theories naturally splits into $W > 1/3$ and $W < 1/3$ subfamilies.  
The $W < 1/3$ family should be qualitatively distinct from a quantum-mechanical perspective, as outlined at the end of the next subsection.  
A particularly simple example of such a dynamically-consistent theory is the $W = 0$ theory, for which the constraints are
\be 
\begin{array}{cc} 
p^{ab}p_{ab} = g\Lambda, & {p^{ij}}_{;j} = 0.
\end{array}
\label{WSConstraints}
\ee
W = 0 may be of particular relevance, in part because conformal gravity has been formulated i.t.o  the $W = 0$ supermetric, and in part from string-theoretic considerations 
(see next subsection).  Conformal gravity arises because, in a conformal generalization of the above working, the equivalent of the slicing equation (\ref{slicingeq}) is 
independently guaranteed to hold.  There is also a separate strong conformal theory \cite {CG}.

Since for $s = 0$ (\ref{Sevolham}) reduces to 
\be
\dot{{\cal H}} = \frac{(3X - 2)Np}{2\sqrt{g}}{\cal H} + \pounds_{\xi}{\cal H},
\ee
the momentum constraint may no longer be seen to arise as an integrability condition.  This fact was already noted by Henneaux \cite{Henneaux}.
Strong gravity thus provides a counterexample to the suggestion that all additional constraints 
need arise from the propagation of ${\cal H}$.  However, all the other constraints can be interpreted as arising in this way in the standard 
approach to GR \cite{Vanderson}: ${\cal H }_i$, the electromagnetic Gauss constraint ${\cal G}$, the Yang--Mills Gauss constraint ${\cal G}_J$ and the `locally Lorentz' 
constraint ${\cal J}_{AB}$ from working in some first order formalism.

\mbox{ }

It is also dynamically consistent \cite{Rovelli} to have strong gravity without a momentum constraint ${\cal H}_i$ to start off with; the new $W \neq 1$ theories may be treated 
in this way too.  In fact, it is this treatment that corresponds to strictly taking the $G \longrightarrow \infty$ limit of GR (as opposed to Pilati's approach \cite{PilatiID}
in which the momentum constraint is kept).  This is because the GR momenta are proportional to $G^{-1}$ \cite{bombelli}.  
So there are distinct strong gravity theories with 5 and 2 degrees of freedom per space point respectively.  
Also, if one uses Ashtekar variables in place of the traditional ones,  the analogues of ${\cal H}_i$ and ${\cal J}_{AB}$ cease to be independent, so one is 
forced to have the theory with 5 degrees of freedom.   
\cite{bombelli}.  

Starting off without a momentum constraint ammounts to starting off with `bare velocities' rather than best-matched ones, which corresponds to another 3-space scheme 
suggested in the GR case by \'{O} Murchadha \cite{Niall}.  Our generalization to arbitrary $s$ of this leads to
\be
\dot{{\cal H}} = -\frac{2s}{N}(N^2[{p^{ab}}_{;b} + (X - 1)p^{;a}])_{;a}
+ \frac{(3X - 2)Np}{2\sqrt{g}}{\cal H}.
\ee
For $s \neq 0$, $ {p^{ab}}_{;b} + (X - 1)p^{;a} = 0$. But propagating this gives 
\be
\begin{array}{ll}
\frac{\pa}{\pa\lambda}[{p^{ab}}_{;b} + (X - 1)p^{;a}] = & 
(X - 1)
\left(
\left[
2s(NR - \nabla^2N) + 3N\Lambda + \frac{N(3X - 2)p^2}{2g}
\right]^{;a} 
- \frac{(3X - 2)Np}{2}
\left(
\frac{p}{\sqrt{g}}
\right)^{;a}
\right)
\\ &
-\frac{1}{2N}(N^2{\cal H})^{;a}
- \frac{2N}{\sqrt{g}}\left(p^{am} - \frac{1}{2}g^{am}p \right){p^b}_{m;b},
\end{array}
\ee
so if constraints alone arise (rather than conditions on $N$), we require $X = 1$ and we recover relativity: the constraint ${\cal H}_a \equiv -2{p_{ab}}^{;b} = 0$ 
may be encoded into the bare action by the introduction of an auxiliary variable $\xi^a$.  
This encoding may be thought of as the content of best matching.  
Clearly, the $s = 0$ example is of value because it illustrates that it is possible to 
`miss out' constraints if we interpret these as integrability conditions for 
${\cal H}$.  Thus although the `bare' and `best-matched' schemes are equivalent for pure GR, they are not in general equivalent.  

\subsection{Application to Scalar-Tensor Theories}

The $X \neq 1$ departure from the DeWitt supermetric does not appear to affect Henneaux's study of the geometry.  Whereas these theories are no longer 
interpretable as truncations of GR, they do correspond to truncations of 
scalar-tensor theories (such as Brans--Dicke theory), in a region where the scalar field is a large constant.  The relations between the Brans--Dicke parameter $\omega$ and 
our coefficients $W$ and $X$ are shown in Fig. 2.  We now discuss the possibility that a positive-definite ($W < \frac{1}{3}$ i.e 
$\omega < 0$) inner product can occur in our universe.   There is no point in considering Brans--Dicke theory since this has $\omega$ constant in space and time and we know 
from solar system tests that today $\omega \geq 3500$ \cite{Will}, corresponding to $W$ being very slightly larger than the GR value 1.  However, general scalar-tensor theory 
permits $\omega$ to vary  with space and in particular time, so it could be that the very early universe had a very different value of $\omega$ from that around us today, since 
the bounds on $\omega$ from nucleosynthesis \cite{dampich} permit $\frac{\omega_{\mbox{\scriptsize nucleosynthesis\normalsize}}}{\omega_{\mbox{\scriptsize today\normalsize}}} 
\approx \frac{1}{25}$.  The bounds from \cite{KiefGiu} are less strict but presumably applicable to a wider range of theories since the origin of the departure from $W \neq 1$ 
(i.e $\omega \neq \infty$) is there unspecified.  Furthermore, $\omega$ is attracted to the GR value at late times in scalar-tensor theories \cite{letter, paper}, so it need not 
have started off large.  One would expect $\omega$ of order unity in any fundamental scalar-tensor theory \cite{letter}.  For example $\omega = -1$ arises in low-energy string 
theory \cite{BD}.    

It is thus an open question whether $\omega$ at early times could have passed through the value 0.  This question is interesting for the quantum-mechanical reasons given 
in the next paragraph.  We first wish to clarify the role of our strong gravities in such a study.  Our strong gravity theories do \it not \normalfont permit $W$ (and hence 
$\omega$) to 
change with time, so we are not suggesting to use these to investigate whether such a transition through $\omega = 0$ is possible.  
But if such transitions are found to  be possible, the very early universe could then be described by scalar-tensor theories which have $\omega < 0$.  Then one of our 
strong gravity theories which behaves qualitatively differently 
from the usual $W = 1$ strong gravity corresponding to GR would be relevant as the approximation near the initial singularity.  We propose to study the possibility of having 
such a transition using the full  scalar-tensor theories.  Unlike their strong gravity limits discussed in this paper, 
for which this transition involves passing through a noninvertible supermetric,\fn{In fact, this $W = \frac{1}{3}$ supermetric is of the same form as the degenerate strong 
gravity 4-metric, so the pointwise geometry of superspace for $W = \frac{1}{3}$ should be taken to be akin to Henneaux's geometry of strong gravity spacetimes.} 
the full scalar-tensor theories are not badly-behaved as $\omega \longrightarrow 0$.  This is because despite the degeneracy of the tensor (`gravity') supermetric 
for $\omega = 1$, what counts for the full theory is the larger\fn{Just like 
the GR supermetric may be represented by a $6 \times 6$ matrix \cite{DeWitt}, 
the scalar-tensor supermetric may be represented by a $7 \times 7$ matrix where 
the new seventh index is due to the scalar field.  For an account of this in Brans--Dicke theory, which also explains how other-signature  
supermetrics can occur elsewhere in full Brans--Dicke theory, see \cite{KM}.} scalar-tensor supermetric, which is usually well-behaved at $\omega = 0$ because of the presence 
of scalar-tensor cross-terms in the scalar-tensor supermetric, due to which the degeneracy of the $6 \times 6$ block is not sufficient to cause the whole $7 \times 7$ 
scalar-tensor supermetric to be degenerate.  But in the approximation by which the theories in this paper arise from scalar-tensor theories the scalar momentum is negligible, 
so one \it is \normalfont then left with only the `gravity' supermetric.     

Thus in principle there could be different possible early universe behaviours which admit 3 different sorts of strong gravity limits: $\omega > 0$ (indefinite), 
$\omega = 0$ (degenerate) and $\omega < 0$ (positive-definite).   In each of these cases the natural inner product corresponding to the early-universe Wheeler--DeWitt 
equation (quantum Hamiltonian constraint) will be very different.    This will cause the early universe's quantum theory of gravity in the $\omega < 0$ (and $\omega = 0$) 
cases to be substantially different from quantum general relativity, which belongs to the $\omega > 0$ case, by the following argument.  Finding an inner product is crucial 
for the physical interpretation of a quantum system.  The natural inner product potentially provides an easy solution to the problem of finding an inner product.  But the 
natural inner product is only an acceptable solution to this problem if it admits a probabilistic interpretation \cite{Kuchar92}.\fn{In particular, the indefinite natural 
inner product of GR does \it not \normalfont admit a Klein--Gordon-type inner product.  This is the inner product problem of quantum GR.}  
Now, the arguments for whether the natural inner product admits a probabilistic interpretation depend on the type of the natural inner product: given a positive-definite inner product, one must check whether it admits a 
Schrodinger-type probabilistic interpretation, which is a considerably different procedure from checking whether a given indefinite inner product admits a Klein--Gordon-type 
probabilistic interpretation.    

The extension to $W \neq 1$ of Isham and Pilati's perturbative quantization idea of expanding about strong gravity  
ought to be both tractable and of relevance to such a quantum study of scalar-tensor theories.  The idea involves expanding about the strong gravity theory by  
introducing a comparatively small $R$ term.   In the GR case which alone has been investigated \cite{Pilatilit}, one thus recovers a GR regime away from the singularity.

\subsection{Application to Conformal Gravity}

Another application of our $W \leq \frac{1}{3}$ theories follows from conformal gravity being $W$-insensitive and thus expressible i.t.o a $W = 0 < \frac{1}{3}$ supermetric.     
In fact, we can form a sequence of theories: $W = 0$ strong gravity, conformal strong gravity, conformal gravity.   
Thus we can isolate the study of some of the novel features of classical and quantum conformal gravity:   
$W = 0$ strong gravity has a positive-definite inner product, and in    
strong conformal gravity additionally ${\cal H}$ begins to play a new role and there is a preferred slicing.  
Conformal gravity has in addition a nontrivial integro-differential lapse-fixing equation to solve (Whereas strong conformal gravity's lapse-fixing equation merely  
integrates to give $N = \mbox{spatial constant}$), which represents an additional computational step before one can attempt to solve the evolution equations.  

In answer to whether arguments from the Sec 2.1 are applicable to conformal gravity, we begin by noting that there is no `expansion term' $p$ in conformal gravity.    
Because this absence is due to $p = 0$ being seperately variationally imposed (rather than due to $W = 0$ occurring for the vacuum theory), 
the presence of non-minimally-coupled-scalars or the related use of conformal transformations are unable to reintroduce a $p$ into conformal gravity.    
A consequence of this absence is that the usual notion of cosmology is not applicable to conformal gravity.  It is not currently known whether any conventional 
or non-conventional cosmology can be recovered from conformal gravity by other means.  Also, because $p = 0$ is seperately imposed, 
conformal gravity cannot be included among the `wider range of theories' for which the less stringent bounds on $W$ mentioned in 2.1 are applicable.  
Conformal gravity is a theory in which $W$ plays no role at all.  Presumably the classical and quantum study of conformal gravity on superspace with 
$W <\frac{1}{3}$ and $W < \frac{1}{3}$ are equivalent once projected down to conformal superspace.   
Working out how this happens may be interesting and instructive, at least from a theoretical point of view.

\subsection{Difficulty with Implementation of Ashtekar Variables}

This paper proposes theories for which the inner product problem of quantum GR is altered (if not ameliorated).  
In the case of conformal gravity, the presence of a  preferred foliation additionally represents an attempt to circumvent the problem of time of quantum GR.  
One must however recall that these problems of quantum gravity are always intertwined with other formidable problems, which include operator ordering and regularization 
\cite{Kuchar92}.  At least in GR, Ashtekar variables \cite{Ashtekar} have nice properties as regards these last two problems: the constraints become polynomial functions (cutting 
down on the ordering ambiguities) and a natural regularization is provided.    It therefore becomes of interest whether scalar-tensor theories or conformal gravity admit an 
analogue of Ashtekar variables.  Indeed, how special is GR in admitting Ashtekar variables with their nice properties?  
The Ashtekar variables for GR \cite{Ashtekar} are an $SU(2)$ connection $A_a^{AB}$ and its conjugate soldering form ${\overline{\sigma}^a}_{AB}$ (which is related to the 3-metric 
by $g_{ab} = - tr(\overline{\sigma}_a\overline{\sigma}_b)$).\fn{The overline denotes that the soldering form is a densitized object: i.e it contains a factor of $\sqrt{g}$.   
The capital indices in this section are spinorial internal $SU(2)$ Yang--Mills indices. 
$tr$ denotes the trace over these. $D_a$ is the $SU(2)$ covariant derivative 
as defined in the first equality of (\ref{ashgauss}), and 
$[\mbox{ },\mbox{ }]$ is the $SU(2)$ commutator.}  
One then has the constraints 
\be
D_a{\overline{\sigma}^a}_{AB} \equiv \pa_a{\overline{\sigma}^a}_{AB} +  [A_a,{\overline{\sigma}^a}]_{AB} = 0 
\label{ashgauss}, 
\ee
\be
tr({\overline{\sigma}^a}F_{ab}) = 0 
\label{ashmom}, 
\ee
\be
tr({\overline{\sigma}^a}{\overline{\sigma}^b}F_{ab}) = 0 
\label{ashham}.
\ee
(\ref{ashgauss}) arises for the same reason as the ${\cal J}_{AB}$ mentioned on page 5: it is a `locally Lorentz' constraint corresponding to the use of a first-order formalism.  
Note that in the Ashtekar formalism, (\ref{ashgauss}) has the form of an $SU(2)$ Yang--Mills--Gauss constraint.  (\ref{ashmom}) and (\ref{ashgauss}) are the polynomial forms that the 
momentum and Hamiltonian constraints respectively, where ${F_{ab}}^{AB} \equiv 2\partial_{[a}{A_{b]}}^{AB} + [A_a, A_b]^{AB}$ is the field strength corresponding to ${A_a}^{AB}$.  One can 
see from a gauge-theoretic point of view that (\ref{ashmom}) is indeed associated with momentum since it is the condition for a vanishing Poynting vector.  The Hamiltonian 
constraint has no such obvious gauge-theoretic interpretation.  

By including a cosmological constant term\fn{In Ashtekar variables, 
$g$ is proportional to $\epsilon_{abc}{\overline{\sigma}^a}{\overline{\sigma}^b}{\overline{\sigma}^c}$, so the cosmological constant term itself is also polynomial.} 
and passing to the strong-coupled limit of GR as above, 
it is simple to show that the constraints (\ref{ashgauss}), (\ref{ashmom}) and (\ref{ashham}) become (see also \cite{bombelli,Husain})  
\be
[A_a, \overline{\sigma}^a] = 0,
\label{sashgauss}
\ee
\be
tr(\overline{\sigma}^m[A_m, A_i]) = 0,
\label{sashmom}
\ee
\be
tr(\overline{\sigma}^a\overline{\sigma}^b[A_a, A_b]) - g\Lambda = 0.  
\label{sashham}
\ee
We can easily see, by the cyclic property of the trace and use of (\ref{sashgauss}),  that (\ref{sashmom}) is redundant as claimed on page 5.  Furthermore, there is an 
equivalent form for the remaining constraints (\ref{sashgauss}) and (\ref{sashham} )  \cite{bombelli}, which we express as 
\be
A_{[ab]} = 0,
\label{equivsashgauss}
\ee
\be
A_{ab}A_{cd}\underline{G}^{abcd} - \Lambda = 0
\label{equivsashham}
\ee
(for $A_{ab} \equiv tr(A_a\sigma_b)$), to manifestly display the dependence on the 
(now overall undensitized inverse) DeWitt supermetric $\underline{G}^{abcd} = \frac{1}{\sqrt{g}}G^{abcd}$.  We then investigate what happens when 
$G^{abcd}$ is replaced by $G_{\mbox{\scriptsize W\normalsize}}^{abcd}$.  Notice how then the Hamiltonian constraint no longer contains a truncation of the natural object 
${F_{ab}}^{AB} = 2\partial_{[a}{A_{b]}}^{AB} + [A_a, A_b]^{AB}$ of $SU(2)$ Yang--Mills theory, in correspondence with $W \neq 1$ strong gravity not being a natural truncation of the GR 
Hamiltonian constraint.  We find that whereas (\ref{equivsashgauss}) and (\ref{equivsashham}) still close for arbitrary $W$, the closure of the original (\ref{sashgauss}) and 
(\ref{sashham}) requires the bracket in (\ref{sashham}) to be antisymmetric, that is $W = 1$.  

For full scalar-tensor theory, we do not think Ashtekar variable with $W \neq 1$ will work.  One has there the option of making conformal transformations to put scalar-tensor 
theory into a $W = 1$ form, but the conformal factor required then causes the constraints to be non-polynomial \cite{Capovilla}.  As for conformal gravity, we could as well 
write the theory with $W = 1$, but we see no way that conformal gravity's analogue of (\ref{slicingeq}) can be expressed polynomially.  
So for all these theories, we cannot so easily imitate the Ashtekar formulation means by which GR can be made to elude the operator ordering problem.

\section{Coupling Matter to Strong Gravity}

We now attempt to couple matter to this theory, following the procedure of BF\'{O}. This enables comparison with the GR case, and leads to a better understanding
of how the 3-space approach works.  

\subsection {Inclusion of Scalar Field}

We include first a single scalar field by considering the action 
\be
S^{(\mbox{\scriptsize strong\normalsize})}_{\mbox{\scriptsize BSW\normalsize}_{\phi}}
= \int d\lambda \int d^3x \sqrt{g} \sqrt{\Lambda + U_{\phi}}
\sqrt{T_{\mbox{\scriptsize W\normalsize}} + T_{\phi}},
\ee
with the gravitationally best matched scalar kinetic term $T_{\phi} = (\dot{\phi} - \pounds_{\xi}\phi)^2$ and a scalar
potential ansatz $U_{\phi} = -(C/4)g^{ab}\phi_{,a}\phi_{,b} + V(\phi)$.  

The conjugate momenta are given by the usual expression (\ref{Sgmom}) and by 
\be
\pi \equiv \frac   {\partial\mbox{\sffamily{L}\normalfont}}
{\partial \dot{\phi}}  = \frac{\sqrt{g}}{2N} ( \dot{\phi}
-\pounds_{\xi}\phi),
\ee
where now $ 2N = \sqrt {(T_{\mbox{\scriptsize W\normalsize}} + T_{\mbox{\scriptsize $\phi$\normalsize}})/ (\Lambda + U_{\mbox{\scriptsize $\phi$\normalsize}})} $.  These can be 
inverted to obtain expressions for $\dot{g}_{ij}$ and $\dot{\phi}$.  The local square root gives the primary Hamiltonian constraint 
\be
^{\phi}{\cal H } \equiv  \frac{1}{\sqrt{g}}\left(p^{ij}p_{ij} - \frac{X}{2}p^2 + \pi^2\right) - \sqrt{g}(\Lambda + U_{\phi}) = 0.
\ee
Variation with respect to ${\xi}_i$ gives the secondary momentum constraint 
\be
^{\phi}{\cal H}_i = -2{p_{ij}}^{;j}  + \pi\phi_{,i} = 0.
\ee

The constraint $^{\phi}{\cal H }$ contains the canonical propagation speed $\sqrt{C}$ of the scalar field.  A priori, this is unrestricted.  However, imposing 
$^{\phi}\dot{{\cal H } } \approx 0$ gives\fn{This has been obtained using a new method that uses the Euler--Lagrange equations implicitly,
so we feel no need to include these cumbersome expressions in this or the next section. This method (similar to that used in \cite{CG}) will be in the author's thesis.}  
\be
\begin{array}{c}
\frac{\partial}{\partial\lambda}\left[\frac{1}{\sqrt{g}}\left(p^{ij}p_{ij} - \frac{X}{2}p^2 + \pi^2\right) - \sqrt{g}(\Lambda  + U_{\phi})\right] =
\frac{Np(3X - 2)}{2\sqrt{g}}
\left[\frac{1}{\sqrt{g}}\left(p^{ij}p_{ij} - \frac{X}{2}p^2 + \pi^2\right) - \sqrt{g}(\Lambda  + U_{\phi})\right] \\
+ \pounds_{\xi}\left[\frac{1}{\sqrt{g}} \left(p^{ij}p_{ij} - \frac{X}{2}p^2 + \pi^2\right) - \sqrt{g}(\Lambda  + U_{\phi})\right]
+ \frac{C}{N}(N^2\pi\phi_{;i})^{;i},
\label{evolham}
\end{array}
\ee

The theory has just one scalar degree of freedom, so if the cofactor of $C$ in the last term were zero, the scalar dynamics would be trivial. Thus one has derived that $C = 0$: 
the scalar field theory cannot have any spatial derivatives.  So, strong gravity necessarily induces the Carroll group structure on scalar fields present, thereby forcing these 
to obey Klauder's ultralocal field theory \cite{Klauderlit}.  This is analogous to how relativity imposes the light-cone structure on scalar fields present in \cite{BOF}.  

We finally note that these results are unaffected by whether one chooses to use the `bare' instead of the `best-matched' formulation.  

\subsection{Inclusion of K interacting vector fields}

\indent We consider a BSW-type action containing the a priori unrestricted vector fields $A_a^I$, $I = 1 $ to $K$,\fn{Capital Latin letters are used in this section 
for tensorial Yang--Mills-type internal indices; we attach no importance to whether these internal indices are raised or lowered, but their order is important in the GR case.}  
\be
S^{(\mbox{\scriptsize strong\normalsize})}
_{\mbox{\scriptsize BSW\normalsize}_{{\mbox{\tiny A\normalsize}}_I}}
= \int d\lambda \int d^3x \sqrt{g} \mbox{{\sffamily L\normalfont}}
(g_{ij}, \dot{g}^{ij}, A_i^I, \dot{A}^i_I, N, N^i)
= \int d\lambda \int d^3x \sqrt{g} \sqrt{\Lambda + U_{\mbox{\scriptsize A\normalsize}_I}}
\sqrt{ T_{\mbox{\scriptsize W\normalsize}} + T_{\mbox{\scriptsize A\normalsize}_I}}.
\ee
We use the most general homogeneous quadratic best matched kinetic term $T_{\mbox{\scriptsize A\normalsize}_I}$, and a general ansatz for the potential term 
$U_{\mbox{\scriptsize A\normalsize}_I}$.  $T_{\mbox{\scriptsize A\normalsize}_I}$ is unambiguously 
\be 
T_{\mbox{\scriptsize A\normalsize}_I} = P_{IJ}g^{ad}( \dot{A}_a^I - \pounds_{\xi}A_a^I)( \dot{A}^J_d - \pounds_{\xi}A^J_d)
\ee
for $P_{IJ}$ without loss of generality a symmetric constant matrix.  
We further assume that $P_{IJ}$ is positive-definite so that the quantum theory of $A^I_a$ has a well-behaved inner product.  
In this case, we can take $ P_{IJ} = \delta_{IJ} $ by rescaling the vector fields.  

The $U_{\mbox{\scriptsize A\normalsize}_I}$ considered here is the most general 
up to first derivatives of $A^I_{a}$, and up to 
four spatial index contractions. In the GR case, the latter was a good assumption because it is equivalent to the necessary naive power-counting requirement for 
the renormalizability of any emergent four-dimensional quantum field theory for $A_{Ia}$ \cite{AB}.\footnote{We shall however see that for strong gravity these `quantum' 
conditions may be dropped.}   Then $U_{\mbox{\scriptsize A\normalsize}_I}$ has the form 
\be
\begin{array}{ll}
U_{\mbox{\scriptsize A\normalsize}_I} = & 
O_{IK}C^{abcd}A^{I}_{a;b}A^K_{c;d} + {B^I}_{JK}\bar{C}^{abcd}A_{Ia;b}A^J_c A^K_d + I_{JKLM}\bar{\bar{C}}^{abcd}A^J_aA^K_bA^L_cA^M_d \\ & +
\frac{1}{\sqrt{g}}\epsilon^{abc}(D_{IK}A^I_{a;b}A^K_c + 
E_{IJK}A^I_aA^J_bA^K_c) + F_Ig^{ab}A^I_{a;b} + 
M_{JK}g^{ab}A^J_aA^K_b,
\end{array}
\ee
where $C^{abcd} = C_1g^{ac}g^{bd} + C_2g^{ad}g^{bc} + C_3g^{ab}g^{cd}$ is a generalized supermetric, and similarly for $\bar{C}$ and $\bar{\bar{C}}$ with distinct coefficients.  
$O_{IJ}$, $B_{IJK}$, $I_{IJKL}$, $D_{IJ}$, $E_{IJK}$, $F_{IJ}$, $M_{IJ}$ are constant arbitrary arrays.  
Without loss of generality $O_{IJ}$, $M_{IJ}$ are symmetric and $E_{IJK}$ is totally antisymmetric.  

The conjugate momenta are given by (\ref{Sgmom}) and 
\be
\pi_I^i \equiv \frac   {\partial\mbox{\sffamily{L}\normalfont}}
{\partial \dot{A^I_i}}  = \frac{\sqrt{g}}{2N} ( \dot{A_I^i}
-\pounds_{\xi}A_I^i),
\label{SAmom}
\ee
where now $2N = \sqrt {(T_{\mbox{\scriptsize W\normalsize}} + T_{\mbox{\scriptsize A\normalsize}_I})/ (\Lambda + U_{\mbox{\scriptsize A\normalsize}_I})}$.  These can be 
inverted to give expressions for $\dot{g}_{ij}$ and $\dot{A}^I_i$.  The local square root gives the primary Hamiltonian constraint, 
\be
^{\mbox{\scriptsize A\normalsize}_I}{\cal H } = 
\frac{1}{\sqrt{g}}\left(p^{ij}p_{ij} - \frac{X}{2}p^2 + \pi^I_i\pi_I^i\right) 
- \sqrt{g}(\Lambda + U_{\mbox{\scriptsize A\normalsize}_I})  = 0.
\ee
We get the secondary momentum constraint by varying with respect to $\xi_i$:
\be
^{\mbox{\scriptsize A\normalsize}_I}{\cal H}_i =
-2{p_{ij}}^{;j} + \pi^{Ic}({A_{Ic}}_{;i}
- {{A_I}_i}_{;c}) - {\pi_I^c}_{;c}A^{I}_i = 0.
\ee

The evolution of the Hamiltonian constraint is then 
\be
\begin{array}{c}
\frac{\partial}{\partial\lambda}
\left[
\frac{1}{\sqrt{g}}
\left(
p^{ij}p_{ij} - \frac{X}{2}p^2 + \pi^I_i\pi_I^i
\right) 
- \sqrt{g}(\Lambda + U_{\mbox{\scriptsize A\normalsize}_I})
\right] 
= \\
\frac{(3X - 2)Np}{2\sqrt{g}}\left[\frac{1}{\sqrt{g}}\left(p^{ij}p_{ij} - \frac{X}{2}p^2 +\pi^I_i\pi_I^i \right) 
- \sqrt{g}(\Lambda + U_{\mbox{\scriptsize A\normalsize}_I})\right]
+ \pounds_{\xi}\left[\frac{1}{\sqrt{g}}\left(p^{ij}p_{ij} - \frac{X}{2} p^2 + \pi^I_i\pi_I^i \right) 
- \sqrt{g}(\Lambda + U_{\mbox{\scriptsize A\normalsize}_I})\right]\\
- \frac{4}{N}O^{IK} \left(C_1(N^2\pi_I^a{A_{Ka}}^{;b})_{;b}
+ C_2(N^2\pi_I^a{A_K^b}_{;a})_{;b}
+ C_3(N^2\pi_I^a{A_K^b}_{;b})_{;a} \right)\\
- \frac{2} {N} \bar{C}^{abcd}{B^I}_{JK}  (N^2 \pi_{Ia}A^J_cA^K_d)_{;b}
+ \frac{2}{N} \epsilon^{abc}D_{IK}(N^2\pi^I_aA^K_c)_{;b} + \frac{2}{N}F^I(N^2\pi_{Ii})^{;i} \\
+ \frac{1}{N} O^{IK}\left[N^2 (p_{ij} - \frac{Xp}{2} g_{ij}) A_{Kb;d} (2A_I^iC^{ajbd} - A_I^aC^{ijbd}) \right]_{;a}  \\
+ \frac{1}{N} {B^I}_{JK} \left[ N^2 (p_{ij} - \frac{Xp}{2} g_{ij} ) A^J_b A^K_d ( 2A_I^i\bar{C}^{ajbd} - A_I^a\bar{C}^{ijbd}) \right]_{;a}  \\
+ \frac{1}{N} F^I \left[ N^2 (p_{ij} - \frac{Xp}{2} g_{ij} )(2A_I^ig^{aj} - A_I^ag^{ij}) \right]_{;a}.
\label{evolSham}
\end{array}
\ee

We demand that $^{\mbox{\scriptsize A\normalsize}_I}\dot{{\cal H}}$ vanishes weakly.  The first two terms vanish weakly by the Hamiltonian constraint,  
leaving us with nine extra terms.  Because we have less than $3K$ vector degrees of freedom to use up, nontriviality dictates that most of these 
extra terms can only vanish strongly, that is by fixing coefficients in the potential ansatz.  
Furthermore, we notice that all contributions to (\ref{evolSham}) are terms in $N^{;a}$ or are partnered by such terms.  
Since further constraints are independent of $N$, these terms in $N^{;a}$ are of the form $(N^{;a}V_{Ja})S^J$, 
and nontriviality dictates that it must be the scalar factors $S$ that vanish.  
We proceed in three steps.   

$1^{\prime}$) The first, second, third, fifth and sixth non-weakly-vanishing terms have no 
nontrivial scalar factors, so we are forced to have $O^{IK}  = \delta^{IK}$, $C_1 = C_2 = C_3 = 0$, $D_{IK} = 0$ and 
$F_I = 0$. The conditions on the $C$'s correspond to the vector fields obeying the local Carroll structure.  

$2^{\prime}$) This automatically implies that the seventh, eighth and ninth terms also vanish.

$3^{\prime}$) The only nontrivial possibility for the vanishing of the fourth term is if $B_{IJK} = 0$, in which case the constraint algebra has been closed.  

It is enlightening to contrast these (primed) steps with their (unprimed) counterparts from the GR case \cite{AB}.

1) is the same as $1^{\prime}$) except that $C_1 = - C_2 = -1/4$, which corresponds 
to the vector fields obeying the local Lorentz light-cone structure 

2) is the same as $2^{\prime}$) except that instead of the automatic vanishing of the eighth term, one is 
forced to take $B_{IJK} = B_{I[JK]}$, which is the start of the imposition of an algebraic structure on the hitherto unknown arrays.  

3) One is now left with $K$ new \it{nontrivial} \normalfont scalar constraints, which happen to form the Yang--Mills Gauss constraint, 
\be
{\cal G}_J \equiv {\pi_J^a}_{;a} - \mbox{\sffamily g\normalfont}
B_{IJK}\pi^I_aA^{Ka} \approx 0,
\ee
where $\mbox{\sffamily g\normalfont} = -4\bar{C}_1 = 4\bar{C}_2$ will become the coupling constant. 
So the algebra is not yet closed; we have the following further steps. 

Propagation of the new constraints requires that the $M_{IJ}$ and $E_{IJK}$ terms are killed off; 
the first of these conditions means that the vector fields are massless.  Furthermore the propagation forces $I_{JKLM} = {B^I}_{JK}B_{ILM}$ and the Jacobi identity  
${B^I}_{JK}B_{ILM} + {B^I}_{JL}B_{IMJ} + {B^I}_{JM}B_{IJL} = 0$.  So in this case, 
one obtains a gauge theory for which $B_{IJK}$ are structure constants.  
Finally, one is forced to have $B_{IJK} = B_{[IJK]}$, which allows one to restrict 
the algebra associated with the gauge group to being the direct sum of compact simple and U(1) subalgebras by use of the standard Gell-Mann--Glashow result \cite{GMG}, 
albeit in a slightly different way from its use in the flat spacetime derivation of Yang--Mills theory.  The new constraint may now be encoded as the variation in a further 
auxiliary variable introduced according to the best matching procedure corresponding to the gauge symmetry.  Thus one arrives at the (3 + 1) decomposition 
of Einstein--Yang--Mills theory.  

\subsection{Discussion of Matter-Coupling Results} 

The results of this section help clarify some aspects of the 3-space approach results for GR.  First, notice also how now that a family of supermetrics is allowed  
the matter dynamics is insensitive to a possible change of supermetric, 
which is encouraging for the coupling of conformal gravity to matter fields.  

Second, we can take further the view that local causal structure and gauge theory are manifestations of the same thing. 
By the BF\'{O} procedure, in the GR case the universal light-cone and gauge theory come together from the $\dot{R}$ term in  
$^{\mbox{\scriptsize A\normalsize}_I}\dot{{\cal H}}$, 
whilst the absence of this in strong gravity ensures that the collapse of the light-cone to the Carrollian line is accompanied by the breakdown of gauge theory: 
there is neither gauge symmetry nor a Gauss law.  In the GR case, the quantum-mechanics-inspired positive-definiteness assumed of the vector field kinetic matrix 
$P_{IJ}$ then turns out to be necessary in the restriction of the choice of gauge group, so there would be a price to pay if one insisted instead on entirely classical 
assumptions.  In the strong gravity case, the absence of emergent gauge structure means that there is no such price to pay for using classical assumptions alone.  Provided that 
$P_{IJ}$ is invertible, the outcome of steps $1^{\prime}$) to $3^{\prime}$) is unaltered.  

We see through what happens in the absence of the Gauss law in this paper that it is specifically this characteristic of the 3-vector theory that kills off the vector field 
mass terms in the 3-space approach, rather than some underlying principle for general matter.  This is a useful first insight into the status of mass in the 3-space approach 
to GR.  It is also easy to demonstrate that the general  derivative-free potential term built out of vector fields persists coupled to strong gravity.    

We emphasize that our result concerning the breakdown of gauge theory is in particular  a result about GR, although it clearly occurs for all our theories and 
the theories they approximate.  As $G \longrightarrow \infty$ such as in the vicinity  of the initial singularity, dynamical consistency dictates that gauge theory breaks down 
in GR.   Gauge interactions become impossible as one approaches such a regime.  This appears not be in accord with the view that gauge interactions persist in 
extreme regimes to form part of a unified theory with gravity, such as in string  theory.  However, little is known about physics in such regimes, so this classical 
GR intuition might not hold.  If string theory can tame such singularities, the circumstances under which gauge theory breaks down according to GR might not occur.  
However, it could even be that string theory breaks down in such a regime, since according to one interpretation, stringy matter could be a phase of some larger 
theory which breaks down in a high-energy phase transition \cite{stringyrefs2}.  
Also, Carrollian regimes might arise in string theory under other circumstances, and exhibit different behaviour from the strong-coupled limit of GR coupled to gauge 
theory, as suggested by the recent Born--Infeld study \cite{Gibbons}.  

\section{Conclusion}

We have found by the 3-space approach a 1-parameter family of consistent theories 
of evolving 3-geometries.  

Whereas for the parameter value $W = 1$  the corresponding theory is well-known 
to be the strong-coupled limit of GR, we have interpreted our other theories 
as similar limits of scalar-tensor theories, and argued that such scalar-tensor theories may describe our universe.    

For the parameter value $W < \frac{1}{3}$, our theories exhibit qualitatively 
different behaviour from GR, but similar to the behaviour of conformal gravity.  
This qualitative difference shows up in the quantum-mechanical regime by offering 
a different resolution to the inner product problem.  

However, we have argued that this feature is unlikely to be simply combinable 
with the use of Ashtekar variables.  

On coupling matter, our theories enforce that this matter is ultralocal, 
just like BF\'{O}'s treatment of matter enforces a local Lorentz light-cone 
structure on the matter.  Whereas BF\'{O}'s treatment enforces gauge theory, 
the collapse of the light-cone structure in the strong-coupled limit is 
accompanied by the breakdown of gauge theory.

\noindent\large \bf Acknowledgements \normalfont \normalsize

EA is supported by PPARC.  We would like to thank Julian Barbour and Malcolm MacCallum for carefully reading versions of this script, and Brendan Foster 
and Niall \'{O} Murchadha for helpful discussions.

\end{document}